\documentclass[a4paper,11pt]{article}
\usepackage{jinstpub} % for details on the use of the package, please see the JINST-author-manual
\usepackage{lineno}
\title{\boldmath The CMS Fast Beam Condition Monitor for HL-LHC}

\author[a]{G. Auzinger}
\author[h]{H. Bakhshiansohi}
\author[a]{A. Dabrowski}
\author[m]{A.G. Delannoy}
\author[i]{A. Dierlamm}
\author[f]{M. Dragicevic}
\author[h]{A. Gholami}
\author[g]{G. Gomez}
\author[c]{M. Guthoff}
\author[a]{M. Haranko}
\author[a]{A. Homna}
\author[o]{M. Jenihhin}
\author[a]{J. Kaplon}
\author[k, a]{O. Karacheban}
\author[b]{B. Korcsm\'aros}
\author[m, a]{W.H. Liu}
\author[l]{A. Lokhovitskiy}
\author[a]{R. Loos}
\author[i]{S. Mallows}
\author[a]{J. Michel}
\author[c]{V. Myronenko}
\author[d]{G. P\'asztor}
\author[a]{M. Pari}
\author[e]{J. Schwandt}
\author[h]{M. Sedghi}
\author[j]{A. Shevelev}
\author[o]{K. Shibin}
\author[e]{G. Steinbrueck}
\author[j]{D. Stickland}
\author[b]{B. Ujvari}
\author[a]{G.J. Wegrzyn}

\affiliation[a]{CERN, European Organization for Nuclear Research, Geneva, Switzerland}
\affiliation[b]{University of Debrecen, Debrecen, Hungary}
\affiliation[c]{Deutsches Elektronen-Synchrotron, Hamburg, Germany}
\affiliation[d]{ELTE E\"otv\"os Loránd University, Budapest, Hungary}
\affiliation[e]{University of Hamburg, Hamburg, Germany}
\affiliation[f]{HEPHY, Wienna, Austria}
\affiliation[g]{Instituto de Física de Cantabria (IFCA), CSIC-Universidad de Cantabria, Santander, Spain}
\affiliation[h]{Isfahan University of Technology, Isfahan, Iran}
\affiliation[i]{Karlsruher Institut fuer Technologie, Karlsruhe, Germany}
\affiliation[j]{Princeton University, Princeton, NJ, USA}
\affiliation[k]{Rutgers, The State University of New Jersey, Piscataway, NJ, USA}
\affiliation[l]{University of Canterbury, Christchurch, New Zealand}
\affiliation[m]{University of Oxford, Oxford, United Kingdom}
\affiliation[n]{University of Tennessee, Knoxville, TN, USA}
\affiliation[o]{Tallinn University of Technology, Tallinn, Estonia}

\emailAdd{delannoy@cern.ch, olena.karacheban@cern.ch, gabriella.pasztor@cern.ch}

\enlargethispage{12pt}

\abstract{The high-luminosity upgrade of the LHC brings unprecedented requirements for real-time and precision bunch-by-bunch online luminosity measurement and beam-induced background monitoring.
A key component of the CMS Beam Radiation, Instrumentation and Luminosity system is a stand-alone luminometer, the Fast Beam Condition Monitor (FBCM), which is fully independent from the CMS central trigger and data acquisition services and able to operate at all times with a triggerless readout.
FBCM utilizes a dedicated front-end application-specific integrated circuit (ASIC) to amplify the signals from CO$_2$-cooled silicon-pad sensors with a timing resolution of a few nanoseconds, which enables the measurement of the beam-induced background.
FBCM uses a modular design with two half-disks of twelve modules at each end of CMS, with four service modules placed close to the outer edge to reduce radiation-induced aging.
The electronics system design adapts several components from the CMS Tracker for power, control and read-out functionalities.
The dedicated FBCM23 ASIC contains six channels and adjustable shaping time to optimize the noise with regards to sensor leakage current.
Each ASIC channel outputs a single binary high-speed asynchronous signal carrying time-of-arrival and time-over-threshold information.
The chip output signal is digitized, encoded and sent via a radiation-hard gigabit transceiver and an optical link to the back-end electronics for analysis.
This paper reports on the updated design of the FBCM detector and the ongoing testing program.}

\begin{document}
\maketitle
\flushbottom

\section{Introduction}
\label{sec:intro}

\subsection{Upgrade of the BRIL system for the HL-LHC}
The high-luminosity upgrade of the Large Hadron Collider (HL-LHC) aims to increase the peak instantaneous luminosity by a factor of $\approx 5-7.5$ with respect to its design value.
This will correspond to an average rate of interactions per bunch crossing (known as \textit{pileup}) of $\langle \text{PU} \rangle\approx 140-200$, in comparison to a peak pileup of $\approx$60 during LHC Run 3.
The \textit{Phase-2} upgrade of the Compact Muon Solenoid (CMS) detector is designed to ensure efficient operation up to an integrated luminosity of 4000 $\text{fb}^{-1}$.

\begin{figure}[htbp]
\centering
\includegraphics[width=0.9\textwidth]{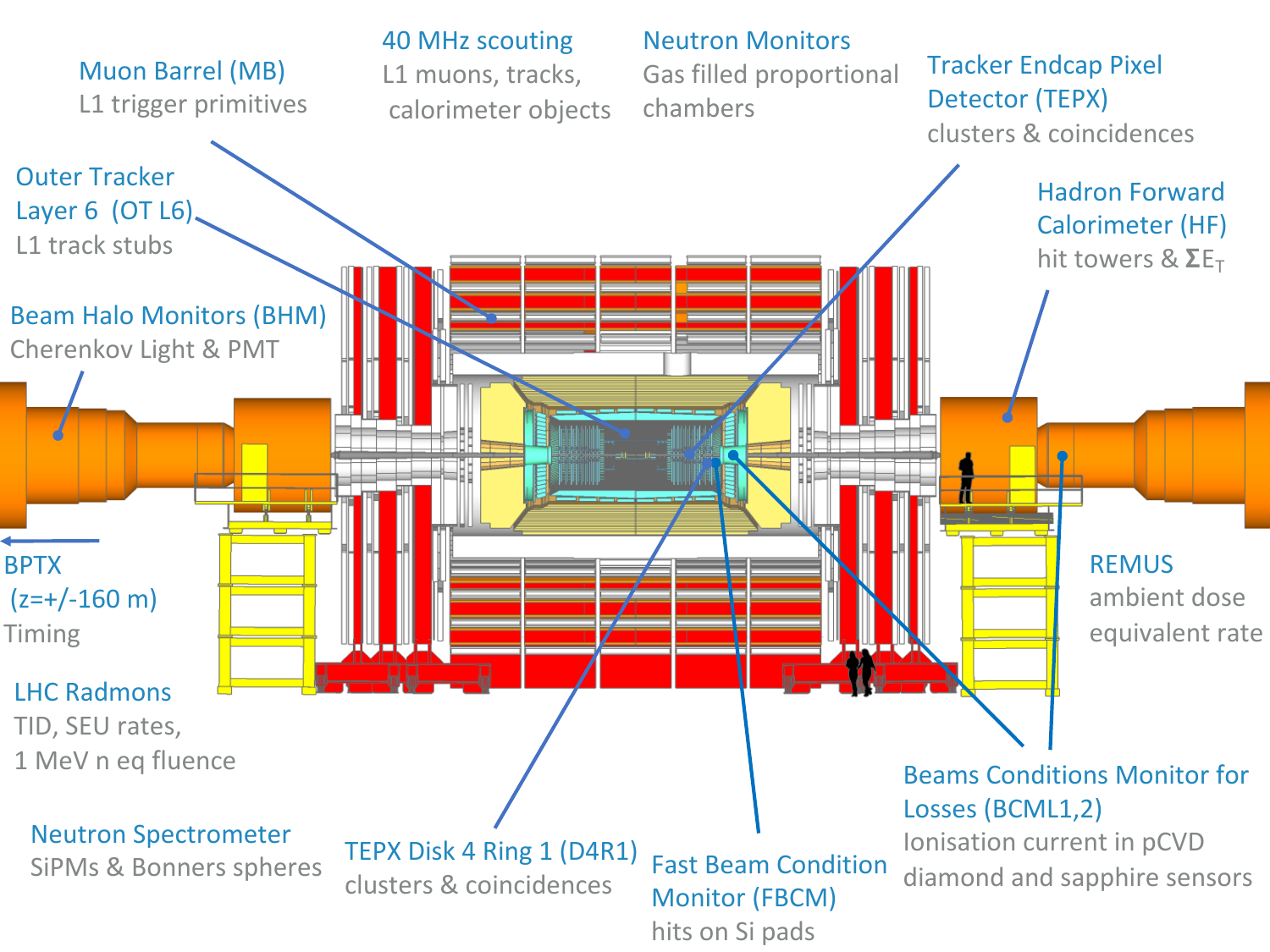}
% https://gitlab.cern.ch/tdr/reports/briltdr2021/-/blob/master/tex/Part1/fig/Intro_HL_LHC/Ph2BRILInstrumentation-2.pdf
\caption{Subsystems of the CMS BRIL project at the HL-LHC.\label{fig:bril_ph2}}
\end{figure}

The Beam Radiation, Instrumentation, and Luminosity (BRIL) project of the CMS experiment is responsible for the operation of multiple detector systems.
Its main deliverables include real-time and precision measurements of the luminosity; monitoring the beam timing, the beam-induced background (BIB) rate, and the beam conditions at the CMS detector to ensure safe operation of the tracker via a beam abort functionality in case of severe beam losses; and finally, monitoring and modelling the radiation environment in the experimental cavern.
The strategy of the BRIL project for Phase-2 combines different elements:
maintenance and upgrade of existing detectors,
development of new instrumentation,
and implementation of dedicated data processing to the backend of other CMS detectors for luminosity determination.
The locations of the Phase-2 BRIL subsystems are shown schematically in Fig.~\ref{fig:bril_ph2}.

\subsection{Requirements for precision luminometry and the dedicated luminosity detector}
\label{subsec:motivation}

% CMS-TDR-023 5.5
The real-time monitoring of the bunch-by-bunch luminosity and BIB is required at all times when beams are circulating.
Luminosity measurement is relied upon to deliver and optimize collisions at each experiment, while the monitoring of the BIB rate is needed to guarantee the safe operation of sensitive subsystems, such as the CMS tracker.
BIB can be measured after a series of unfilled bunches, typically just before a train or in the abort gap, and is expected to arrive to the FBCM location 19 ns before the collision products~\cite{BRIL-TDR}.
Furthermore, special LHC operating conditions (e.g. accelerator commissioning and development, as well as certain periods during the machine cycle before stable beams) require the publication of online luminosity and BIB rate even when the rest of the CMS subsystems may not be in operation.
In order to maximize the availability of these quantities, CMS must be equipped with a dedicated instrument which is independent in its infrastructure and operation from other CMS systems.
Such a detector must be able to operate continuously regardless of the status of other subsystems, and of the central trigger and data acquisition systems.
This independence also ensures flexibility to apply operational changes, such as calibrations, threshold adjustments, high-voltage adjustments, etc. for the optimal performance of the system.

\begin{figure}[htbp]
\centering
\includegraphics[width=1.\textwidth]{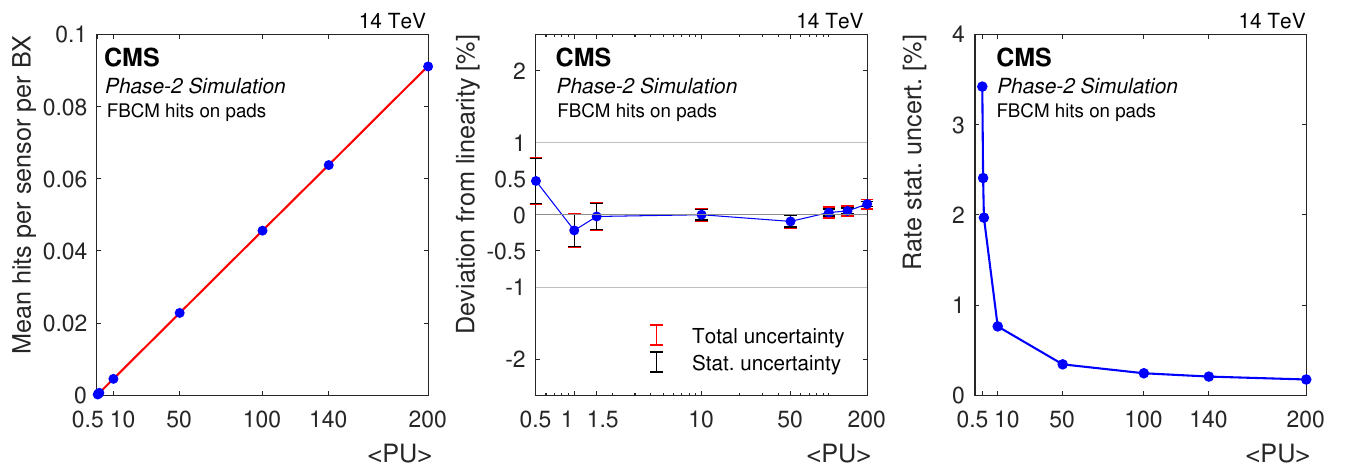}
% https://gitlab.cern.ch/tdr/reports/briltdr2021/-/blob/master/tex/Part2/fig/Fbcm/Residu_Lin_RateErr_vs_pu_v2_final.pdf
\caption{The mean number of hits per sensor per colliding bunch pair in the FBCM luminometer (left), the deviation of the mean hit rate from linearity (center), and the statistical uncertainty in the rate with an integration period of 1 second (right) as a function of pileup. \label{fig:fbcm_uncertainty}}
\end{figure}

% CMS-TDR-023 18.1
% A standalone luminometer able to operate continuously and provide bunch-by-bunch luminosity measurements, independent from the central trigger and data acquisition services, is highly motivated.
An ideal luminometer is expected to show a linear response over the required dynamic range of pileup from $\langle \text{PU} \rangle \sim 0.5$ up to 200 in Phase-2.
Such an instrument must report sufficient bunch-by-bunch rates to achieve sub-percent statistical precision during low-pileup running conditions that arise in special fills for absolute luminosity calibration.
This is a crucial feature since luminosity detectors at hadron colliders must be independently calibrated via the van der Meer (vdM) method~\cite{LUMI-15-16}, where the effective beam overlap is measured from the detector response as a function of the transverse beam separation in special low-pileup conditions tailored to minimize beam -- beam effects.
%A luminosity detector is expected provide a statistical accuracy of $\approx$0.1\% or better on the extracted calibration constant, known as the visible cross section ($\sigma_{\text{vis}}$). -- Olena, I would not go into sigma vis., it's enough to provide the reference for the vdM calibration method. 
%In the context of the HL-LHC, a linear response is required for pileup conditions ranging from 
%<0.1$\times10^{-3}$ $\sim 0.5$to 200 in proton-proton collisions.This requires a detector with small dynamic inefficiencies up to the maximum expected pileup, which reduces nonlinear detector effects.
An optimal luminosity detector must also offer stable long-term performance over the data-taking period, which is monitored and quantified in terms of efficiency and linearity via the analysis of short vdM-like (so-called \textit{emittance}) scans performed during nominal pileup conditions~\cite{EMITTANCE-SCAN}.

The Fast Beam Condition Monitor (FBCM), based on silicon-pad sensors with a fast front-end chip, was proposed as a dedicated luminometer for LHC Phase-2.
The detector concept was described in detail in the Phase-2 BRIL technical design report (TDR)~\cite{BRIL-TDR}.
Based on the simulations of the detector performance, the area and the position of the sensors were optimized.
The optimal location for a sensor of size 2.89 mm$^2$ was found to be at $r\approx14.5~\text{cm}$.
The variation in the mean number of hits per sensor per colliding bunch pair as a function of pileup based on simulation is depicted in Fig.~\ref{fig:fbcm_uncertainty}~(left).
The deviations from linearity are then shown in  Fig.~\ref{fig:fbcm_uncertainty}~(middle), featuring deviations from linearity below $\pm$0.5\%.
The statistical uncertainty of the rate per second is also given in Fig.~\ref{fig:fbcm_uncertainty}~(right), which shows that a statistical uncertainty on the rate of 0.2\% is reached for the physics range of $140 < \langle \text{PU} \rangle < 200$.
These simulations were done for 290 $\mu$m thick Si sensors.
A new round of simulations is ongoing using the updated ASIC design~\cite{FBCM-ASIC}, including the option of thinner 150 $\mu$m sensors, the alternative choice for FBCM (see in more detail in subsection \ref{subsec:sensors}).
\section{FBCM design and readout}
\label{sec:design}

\begin{figure}[htbp]
\centering
\includegraphics[width=.43\textwidth]{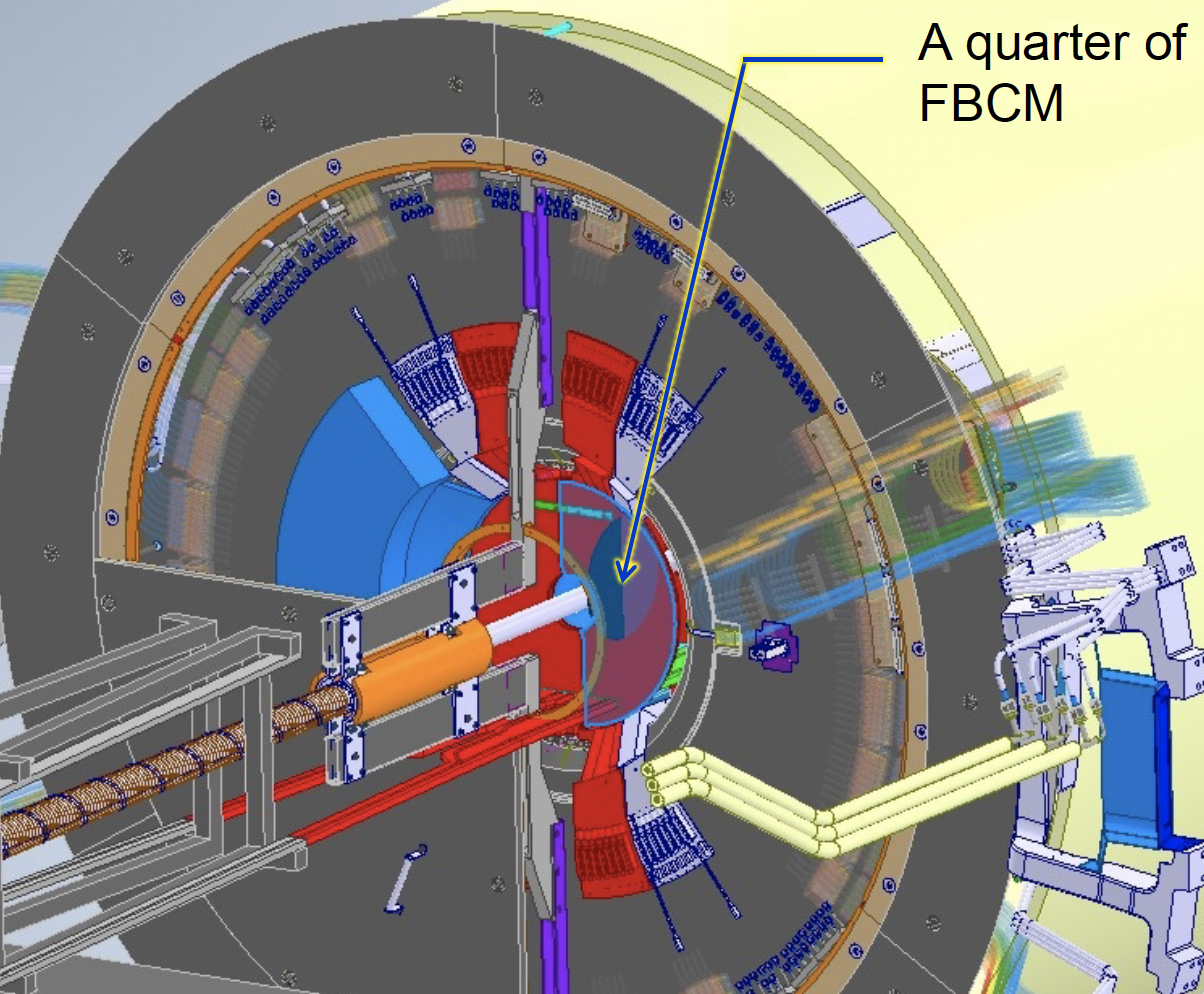} 
\qquad
\includegraphics[width=.4\textwidth]{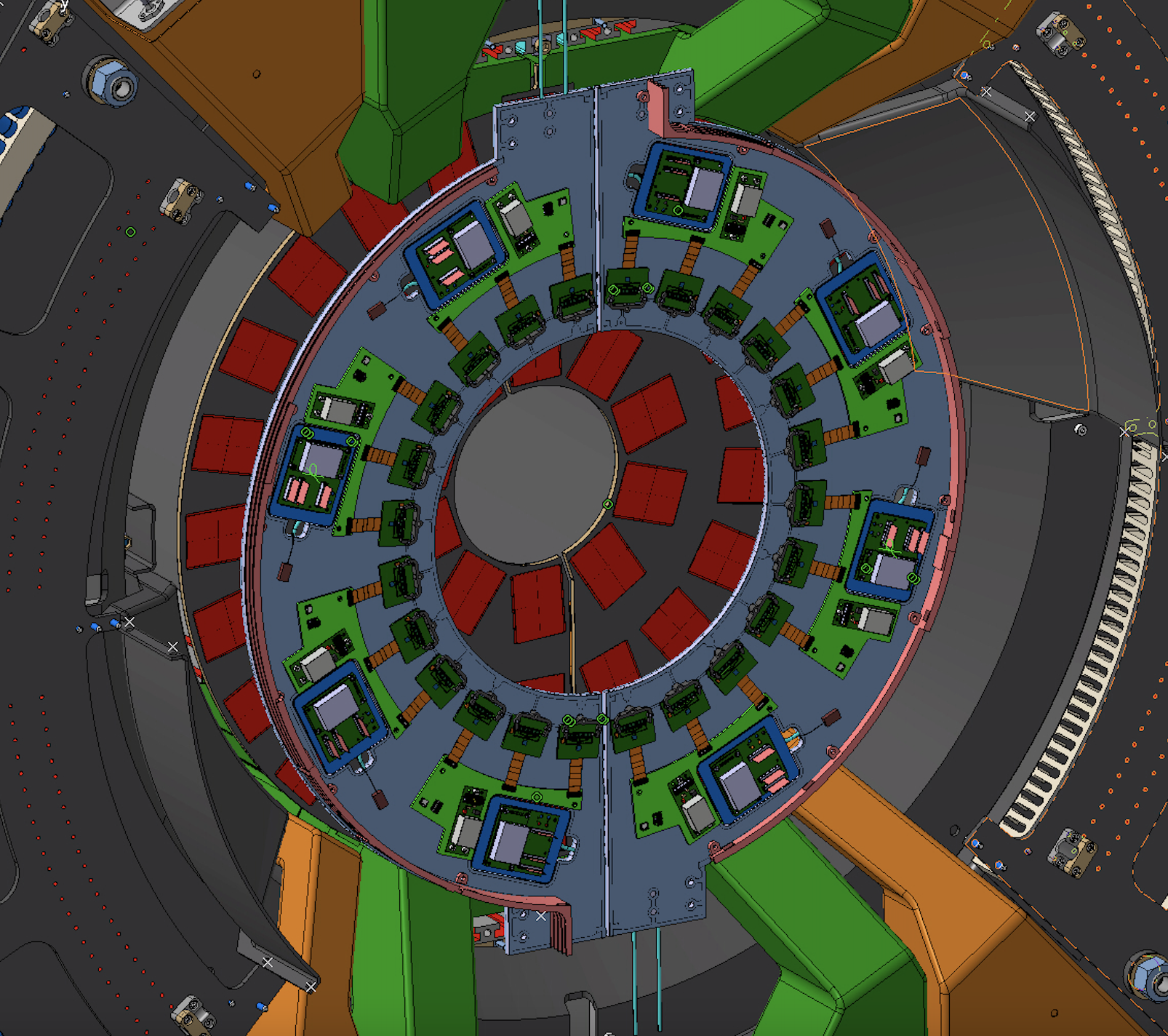}
\caption{Sketch of the FBCM location behind the last disk of the inner tracker (left) and a zoom to the two FBCM half discs forming a ring around the beam pipe with a detailed representation of the components on each service quadrant (right).
\label{fig:FbcmLocation}}
\end{figure}

\begin{figure}[htbp]
\centering
\includegraphics[width=.9\textwidth]{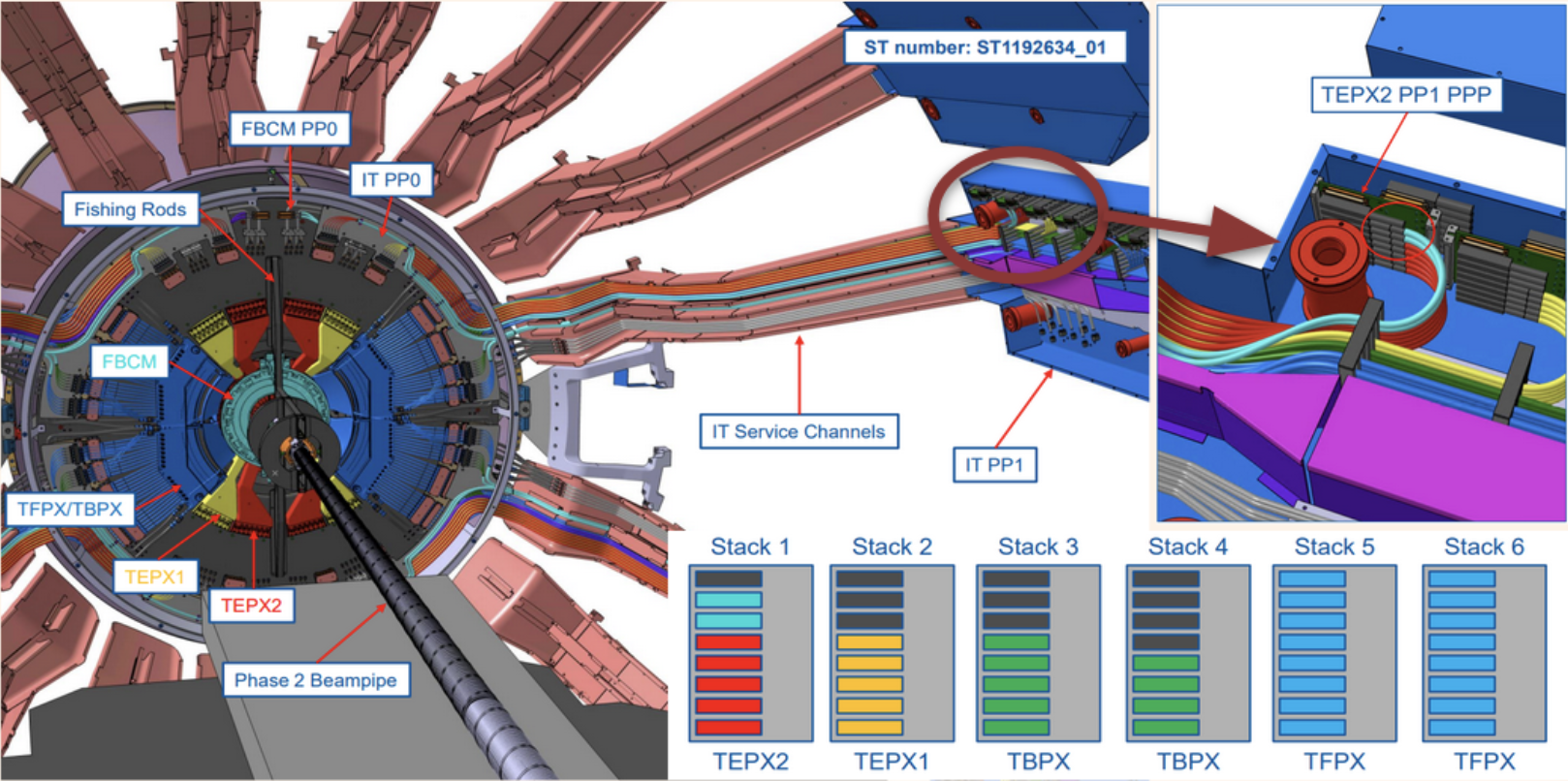}

\caption{FBCM half-disk installation location and position of the patch panels to route the services.
\label{fig:fbcm_pp0_pp1}}
\end{figure}

The FBCM detector concept is based on the BCM1F detector, operated by BRIL during Run 2 and Run 3 of the LHC to provide real-time measurement of luminosity and BIB.
During Run 2, BCM1F was based on a mixture of sensor technologies: in early 2017 ten silicon, ten poly-crystalline diamond (pCVD), and four single-crystal diamond (sCVD) sensors were installed~\cite{BCM1F-diamond, BCM1F-diamond-2}.
%Due to stability and linearity issues, one silicon sensors was tested without cooling.
Since better linearity was observed for the Si-pad sensors, the sensor choice for Run 3 was changed to an all-silicon configuration with a total of 48 channels and active C$_{6}$F$_{14}$ cooling at $-18^{\circ}$C, which led to improved long-term stability and linear response~\cite{BCM1F-RUN3}.
% with increasing integrated fluence and an improved linear response of the sensors as a function of the instantaneous luminosity.
%The BCM1F subsystem consists of four ``C-shaped'' printed circuit boards (PCB) which form two rings on either side of the CMS interaction point.Six double-diode silicon sensors are mounted on each ``C-shaped'' PCB to measure bunch-by-bunch luminosity, using the zero-counting method, and the relative contribution from beam-induced background, determined from the first bunch in the train or a non-colliding bunch.

The FBCM is the next generation of silicon-pad-based luminometer with symmetric and modular design, which avoids single points of failure and simplifies construction and maintenance.
It features a new dedicated front-end ASIC~\cite{FBCM-ASIC}, HL-LHC standard back-end electronics providing fast data processing in FPGA, and active CO$_{2}$  cooling at $-35^{\circ}$C.
%(from C$_{6}$F$_{14}$ at -18$^{\circ}$C to CO$_{2}$ cooling at -35$^{\circ}$C), 
The statistical precision is improved by installing 288 Si-pads.
%additional channels (from 48 in BCM1F to 288 in FBCM), and fast processing in FPGA.
FBCM will be located behind the last disk of the \textit{tracker endcap pixel} (TEPX) detector~\cite{TRACKER-TDR}, close to the bulkhead, as illustrated in Fig.~\ref{fig:FbcmLocation} (left).
The cooling will be connected to the TEPX manifold.

A minimal number of connections per half-disk is foreseen to simplify the installation.
Each half-disk is attached mechanically with two screws.
Each has a single optical connector, and all power connections are provided using two multiservice cables developed for the tracker upgrade.
The FBCM cables will be routed, along with TEPX cables, 
%are reserved in stack 1, 
as shown in turquoise in Fig. \ref{fig:fbcm_pp0_pp1} at stack 1.
A custom patch panel (PP0) will be designed to route all power lines from the inner tracker patch panel (PP1) to FBCM.
The powering scheme was outlined in the technical design report~\cite{BRIL-TDR} and will be finalized in 2024 with the production of the first disk prototype.
Multiple changes were introduced for the mechanical design of the detector to make space for the optical fiber routing and to optimize the thermal contacts.

\begin{figure}[htbp]
\centering
\includegraphics[width=.9\textwidth]{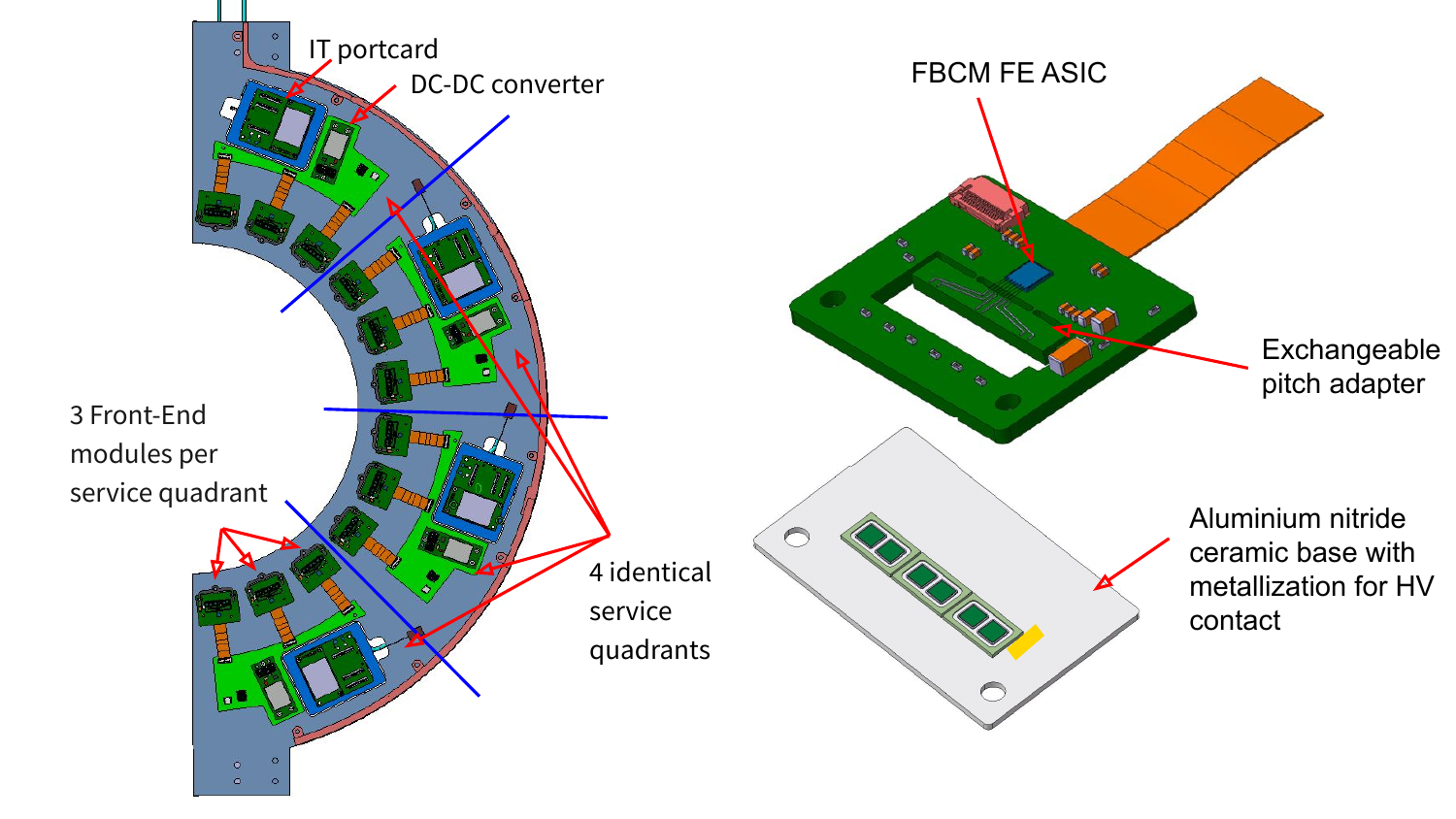}
%https://docs.google.com/presentation/d/1tolMcObXvv7IYcKBer2T2NPyVan81oHPylOUNaNWnOQ/edit#slide=id.g2a153bf5cd2_0_0
\caption{FBCM half-disk (left), as well as the FBCM front-end module and Aluminium-Nitride ceramic base-plate with three two-pad silicon sensors (right).
\label{fig:fbcm_half_disk}}
\end{figure}

The detector is segmented into four mechanically-identical and independent half-disks with inner radius of 8 cm and outer radius of 30 cm.
Two half-disks form a ring around the beam pipe (Fig.~\ref{fig:FbcmLocation} (right)) about 280 cm away from the interaction point on both ends of CMS.
Each half-disk is composed of four identical \textit{service quadrants}, as indicated in Fig. \ref{fig:fbcm_half_disk} (left).
Each service quadrant has an inner tracker (IT) portcard~\cite{Orfanelli:2022zhe}, a bPol12V DC-DC converter~\cite{Faccio:2020rae} (12 V $\rightarrow$ 1.25 V), and a service board~\cite{TRACKER-TDR}, which routes the low voltage and high voltage lines to power three front-end (FE) modules.
Each FE module consists of a hybrid printed circuit board (PCB) housing the ASIC that reads out the  signal of six silicon pad sensors.
The 3x3 mm$^2$ front-end ASIC is equipped with a fast amplifier and a comparator and produces an analog pulse whose rising edge defines the \textit{time of arrival} (ToA) and its duration, defined as the \textit{time over threshold} (ToT), provides information on the deposited charge.
The readout is triggerless.
The signals are routed to the IT portcard \textit{low-power gigabit transceiver}~\cite{Biereigel:2020jri} (lpGBT) which continuously samples 
%the output pulse 
and transmits it to the back-end over an optical link via the \textit{versatile link plus transceivers}~\cite{Soos:2017stv} (VTRx+).
The lpGBT sampling time interval can be as low as 0.78 ns, providing 32 samples per bunch crossing.

The FE hybrid consists of a rigid PCB with an incorporated pitch adapter matching the sensor pads to those of the ASIC, an opening for the sensors, and a flex tail which routes power connections from the service board to the FE hybrid (Fig.~\ref{fig:fbcm_half_disk} (right)).
Silicon sensors are attached with conductive glue to the Aluminium-Nitride (AlN) ceramic baseplate.
Using alignment holes placed in the corners of the hybrid and the baseplate, the sensors are positioned in the opening of the hybrid
%under the rigid part of the FE module PCB 
and then bonded to the ground and, via the pitch adapter, to the ASIC channels.
The surface area of the AlN plate is $30\times20$ mm$^2$ and it provides cooling contact under the sensors and via the hybrid to the ASIC.
An 18 $\mu$m thick layer of copper metalization with golden finish is applied to the AlN ceramic baseplate  to provide high voltage via a wire-bond from the FE board PCB to the sensor back-plane.
The AlN material choice is motivated in the subsection \ref{subsec:cooling}.
The ASIC is wire bonded on the FE module PCB, its position on the PCB is shown in Fig. \ref{fig:fbcm_half_disk} (left).
The ASIC is designed to withstand 200 MRad of total ionizing dose.
The noise from the ASIC is below 900 e$^{-}$.
The first prototype of the FBCM23 ASIC is currently mounted on a testboard that also serves as the prototype of the FE hybrid.
Since its arrival in September 2023, the ASIC undergoes extensive testing~\cite{FBCM-ASIC}.
Irradiation tests, including total ionizing dose (TID) and Single Event Upset (SEU) tests, are in the pipeline.

The \textit{Apollo} board~\cite{Apollo} with two powerful FPGAs, developed for the ATCA crate standard, is chosen for FBCM back-end data processing.
The digital processing unit in the FPGA determines the rising edge and pulse duration, and it aggregates the histograms of the number of hits per bunch crossing identification number (BCID).

\subsection{Sensor design options}
\label{subsec:sensors}

\begin{figure}[!b]
\centering
\includegraphics[width=.8\textwidth]{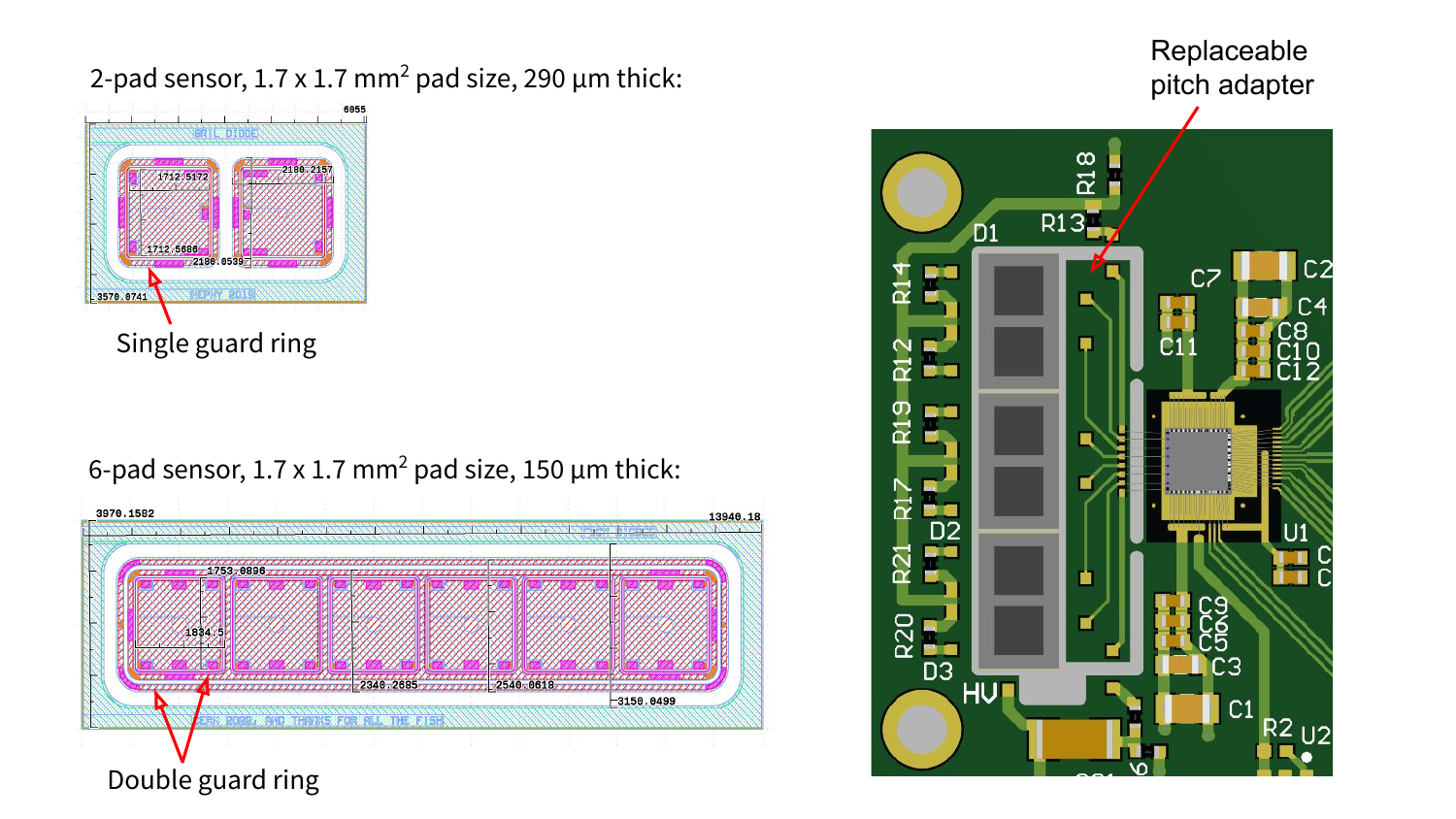}
% https://docs.google.com/presentation/d/1LG4W3Ek8V42APGAFBQNBHdJ2l5YBusRi1j30ELghJX0/edit?usp=sharing
\caption{The two sensor designs for FBCM: a two-pad sensor used in the Run 3 BCM1F (top left), and a new six-pad sensor with double guard ring structure (bottom left), and the FE module prototype with replaceable pitch adapter (right).
\label{fig:fbcm_sensors}}
\end{figure}

Two types of sensors are considered for the FBCM, shown in Fig.~\ref{fig:fbcm_sensors}:
a two-pad, 290 $\mu$m thick sensor which is  used in the Run 3 BCM1F system and manufactured on the half-moons of the CMS Phase-2 outer tracker wafers;
and a six-pad, 150 $\mu$m thick sensor printed on the half-moons of the CMS Phase-2 inner tracker wafers.

At the FBCM sensor location, the expected 1 MeV neutron equivalent fluence is about 2.5 $\times 10^{15}$ cm$^{-2}$ for 3000 fb$^{-1}$ (TID of 200 MRad).
Due to the expected harsh radiation environment, a replacement of the FBCM is planned after about 1500 fb$^{-1}$ (TID of 100 MRad).
Beyond this fluence, the two-pad sensors are not expected to be operational with the required performance.
For comparison, at the outer tracker location, the expected fluence is about $1.5 \times 10^{15}$ cm$^{-2}$, and at  the inner tracker location, about $1 \times 10^{16}$ cm$^{-2}$.
The thicker sensors have better signal-to-noise (S/N) ratio (pre-irradiation S/N = 35; after $1 \times 10^{15}$  cm$^{-2}$ irradiation S/N = 10.7), but larger leakage current.
The thinner sensors are more radiation tolerant, have smaller leakage current, but have smaller signal-to-noise ratio (pre-irradiation S/N = 17; after $1.5 \times 10^{15}$ cm$^{-2}$ irradiation S/N = 8).

The single-pad size for both types of the sensors is about $1.7 \times 1.7$ mm$^2$.
There is a difference of a few mm in the length of the six-sensors panel, depending on the sensor choice.
Using three two-pad sensors results in a longer panel due to sensor edges.
% of the 6 pads on the FE hybrid.
Another significant difference arises from the guard ring structure as the new six-pad sensor features a double ring to improve its performance.
The replaceable pitch adapter (see Fig.~\ref{fig:fbcm_sensors} (right)) allows both types of sensors to be used in the final assembly.
Pitch adapters with several different bonding pad sizes were also produced to evaluate the effect of the parasitic capacitance from the pitch adapter on the signal.
The final sensor choice will be decided in 2024 after a full set of laboratory tests and beam tests with the ASIC and both types of sensors after irradiation.
Both sensor types are available in large quantity.

\subsection{Thermal optimization}
\label{subsec:cooling}

The longevity of FBCM system strongly depends on the efficient cooling of the sensors.
Better thermal contact reduces the dark currents in the silicon sensors and, therefore, extends the lifetime of the detector before a replacement is required.
This is especially important at total ionizing doses approaching 100 MRad.
The FE hybrid design is thermally optimized by incorporating a 0.38 mm thick AlN ceramic baseplate with excellent thermal conductivity of about 180 W/mK, as shown in Fig. \ref{fig:fbcm_half_disk} (right).
As a thermal interface between the AlN and the pocofoam (carbon foam) block, the radiation hard MORESCO RG-42R-1 grease~\cite{FBCM-grease} with 70\% diamond powder will be used to improve thermal contact.
A cooling pipe is placed at the radius of the silicon sensors location and on the return path under the portcard and the bPol12V DC-DC.
Pocofoam blocks (with thermal conductivity of 45 W/mK in plane and 145 W/mK out of plane) will be placed around the cooling pipe to provide thermal contact for AlN under the sensors and as a cooling pad under the bPol12V DC-DC converters.
%Pocofoam thermal conductivity is 45 W/mK in plane and 145 W/mK out of plane.

\begin{figure}[htbp]
\centering
\includegraphics[width=0.9\textwidth]{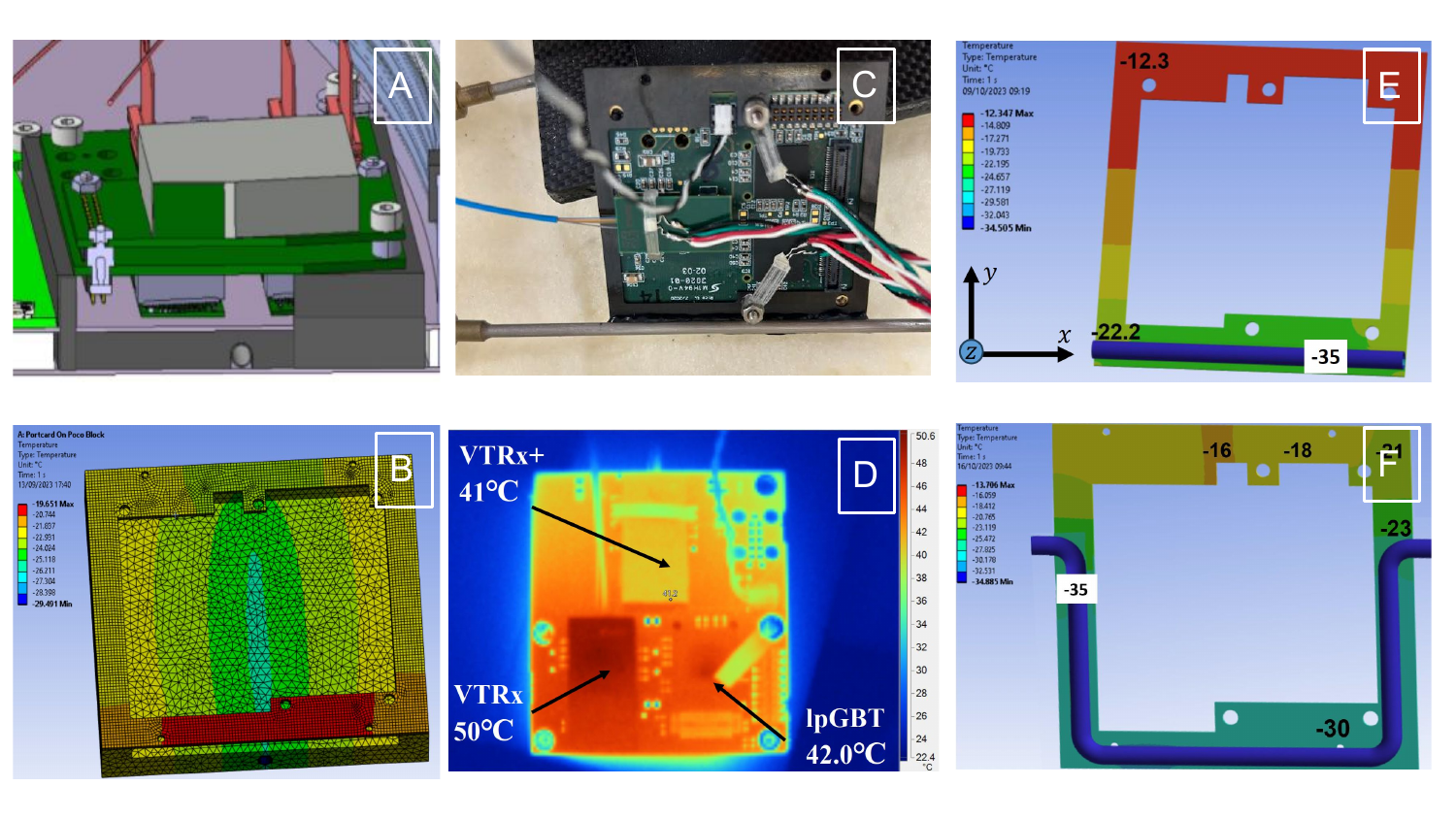}
% https://docs.google.com/presentation/d/1TvPNiqhC-zKBXmpl6c12WgY2hhmXEEEFdBXXxnozZOE/edit?usp=sharing
\caption{A portcard mounted on the FBCM half-disk with a pocofoam insert providing the thermal contact (A).
Comparison of several options validated in ANSYS simulation for the connection of the cooling pipe and the carbon frame (B, E, F).
The setup with portcard mounted on the cooling frame that is attached to a straight cooling pipe and thermal sensors (C).
Thermal camera measurement of a powered portcard (D).
\label{fig:fbcm_portcard_ANSYS}}
\end{figure}

\begin{figure}[htbp]
\centering
\includegraphics[width=0.9\textwidth]{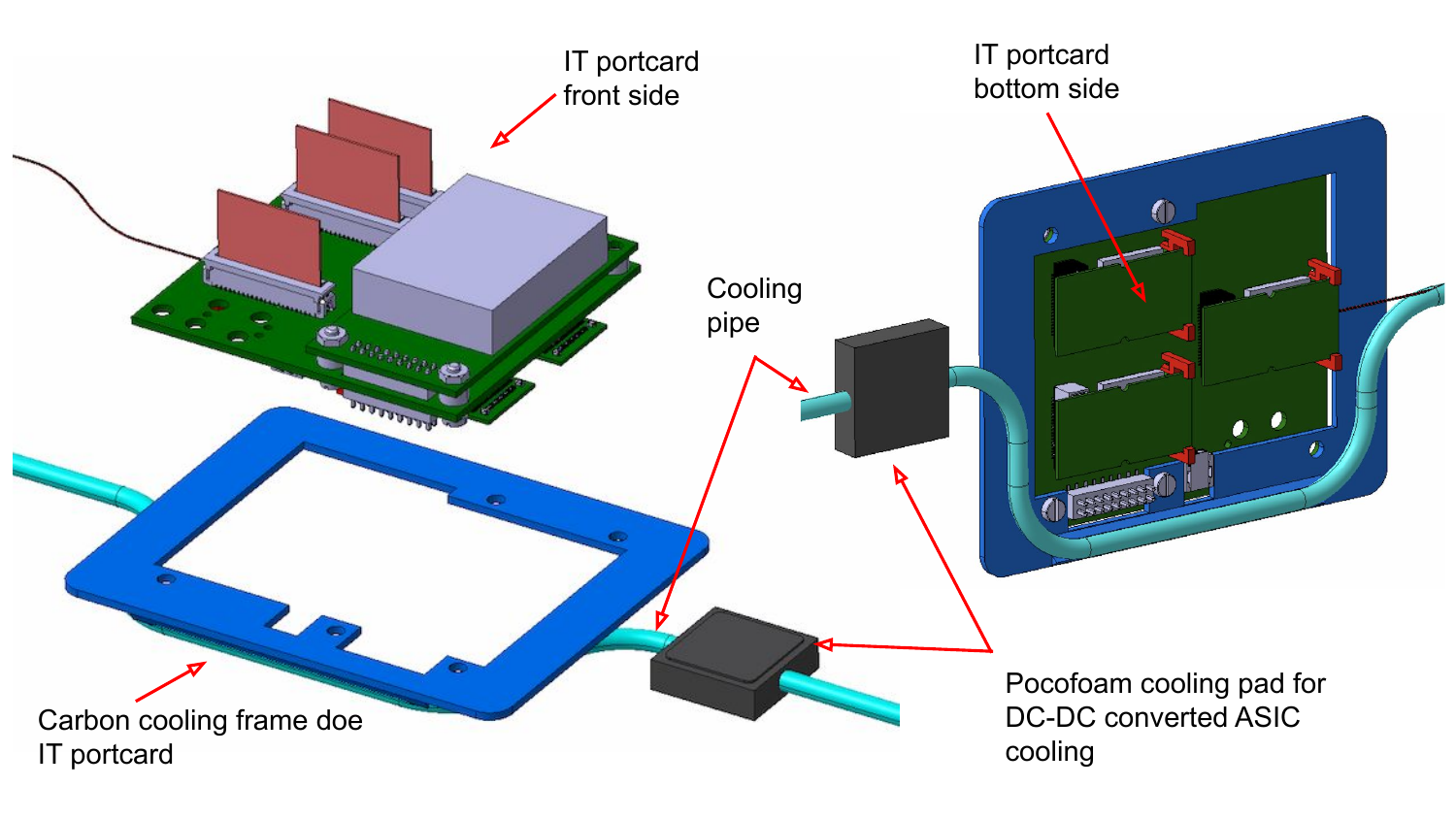}
% https://docs.google.com/presentation/d/1zDsDkOnDTDynInfYSGe_7TOi4hk8IIBDVvrc-wkv2-w/edit?usp=sharing
\caption{Optimized mounting of the portcard using a carbon frame directly in contact with a bent cooling pipe underneath.
\label{fig:fbcm_portcard}}
\end{figure}

Several alternative design options were evaluated for the portcard cooling, as shown in Fig.~\ref{fig:fbcm_portcard_ANSYS}.
They include a straight cooling pipe routed through the pocofoam block and carbon frame on top of it.
The drawing of this design is shown in Fig.~\ref{fig:fbcm_portcard_ANSYS} A, and the corresponding thermal simulation in Fig.~\ref{fig:fbcm_portcard_ANSYS} B.
From the thermal simulation, it is apparent that one side of the frame is not well cooled.
The simulations of the portcard cooling solutions were performed using \textit{ANSYS Mechanical Workbench}.
Another evaluated option was to attach one edge of the carbon frame directly to the cooling pipe.
The picture of the test setup is shown in Fig.~\ref{fig:fbcm_portcard_ANSYS} C, and the thermal simulation on Fig.~\ref{fig:fbcm_portcard_ANSYS} E.
This option only provides good cooling on the side of contact.
Fig.~\ref{fig:fbcm_portcard_ANSYS} D illustrates the location of the heat sources on the portcard.
Finally, a bent cooling pipe that follows half of the contour of the carbon frame, as shown in Fig.~\ref{fig:fbcm_portcard_ANSYS} F, provides the best cooling performance.

The updated detector design with the portcard cooling frame and a bent pipe is shown from the front and back sides in Fig.~\ref{fig:fbcm_portcard}.
The cooling pipe is glued  to the carbon frame with diamond-doped epoxy.
Simulation results agree well with the measurements using PT1000 thermal sensors~\cite{PT1000} and a thermal camera.
The setup was closed in a hermetic box with a flow of dry air.
Thermal camera measurements were only possible at room temperature, when the lid of the box was open.
Therefore, measurements by the PT1000 thermal sensors were first compared at room temperature to the 
thermal camera heat map.
A thermal camera measurement without cooling is shown in Fig.~\ref{fig:fbcm_portcard_ANSYS} D, where it is visible that the VTRx+ at ${50}^\circ$ C and the lpGBT at ${42}^\circ$ C are the hottest components.
With cooling at ${-10}^\circ$ C, the temperature of all components was reduced to below ${20}^\circ$ C.
The cooling performance of the Phase-2 detector is foreseen to be significantly better at ${-35}^\circ$ C.
% The cooling performance will be significantly better at ${-35}^\circ$ C, as foreseen for the cooling system of the Phase-2 detector.
\section{Assembly procedure of FBCM half-disk}
\label{sec:assembly}
% Latest FBCM mechanics and integration photos (FBCM_mechanics_pictures_P2UG_Nov2023.zip, FBCM_intergation_10Nov2023.zip): https://edms.cern.ch/ui/#!master/navigator/document?P:1801174233:101389796:subDocs

\begin{figure}[htbp]
\centering
\includegraphics[width=0.9\textwidth]{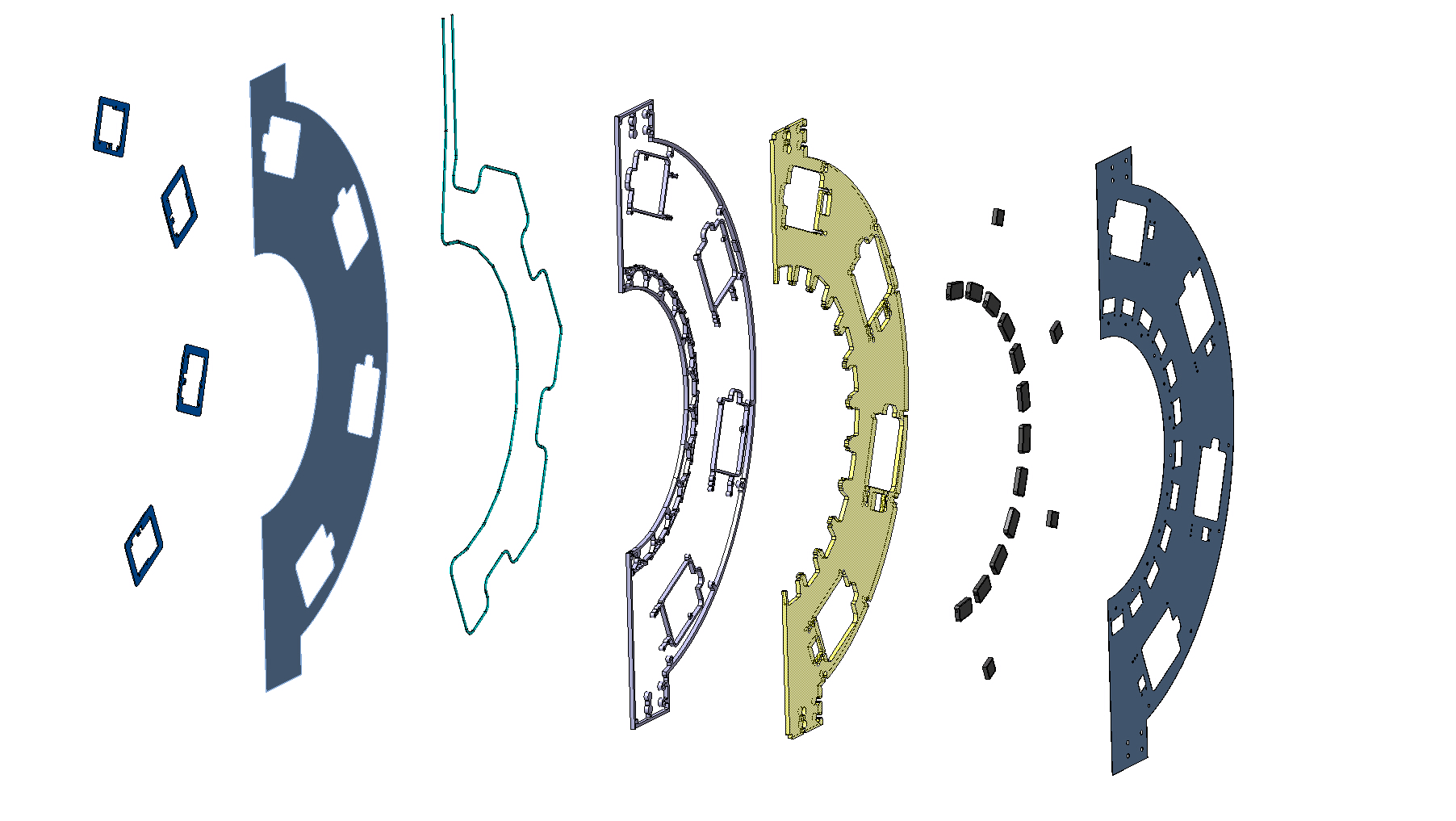}\vspace*{-20pt}
% https://edms.cern.ch/ui/#!master/navigator/document?P:1801174233:101389796:subDocs (FBCM_mechanics_pictures_P2UG_Nov2023.zip)

\caption{From right to left, the FBCM half-disk integration process step by step.
\label{fig:fbcm_building_process}}
\end{figure}

Fig.~\ref{fig:fbcm_building_process} illustrates the FBCM integration process.
The integration stages are shown from right to left.
First, the top carbon fiber sheet is placed on a vacuum plate.
Next, the pocofoam cooling pads under the sensor and DC-DC converter locations are fitted into the cut-outs and they are glued under vacuum, together with the airex foam as a spacer.
Then, the groove for the cooling pipe is machined and the cooling pipe is inserted into the groove using a thin layer of thermally-conductive diamond-doped epoxy.
Finally, the assembly is flattened with a vacuum bag, the bottom carbon fiber sheet is glued on top, and the carbon cooling frames for the portcards are glued on the cooling pipe.

\section{Summary}

The Fast Beam Condition Monitor is a dedicated luminosity and beam-induced background monitor based on silicon-pad sensors with a fast front-end designed for the Phase-2 upgrade of the CMS detector.
It is expected to operate continuously and independently from the central CMS data acquisition and trigger systems, and regardless of the status of other CMS subsystems.
It is designed to have a linear response, from the low pileup conditions during calibration fills, up to the very large pileup values expected in HL-LHC physics conditions.

The FBCM has a symmetric and modular design to avoid single points of failure and simplify maintenance.
%It is segmented into four half-disks, each with four identical service quadrants.Each service quadrant is populated with a service board and three front-end modules.
The design of the mechanics and cooling has significantly matured since the TDR~\cite{BRIL-TDR}, with several optimizations to simplify the integration and installation, and simulation studies to deliver the best cooling performance.

Six silicon-pad sensors are read out by an ASIC, designed for the FBCM, in each front-end module.
The first prototype of the ASIC is undergoing detailed tests at CERN.
% The ASIC produces an analog pulse whose rising edge corresponds to the time of arrival (ToA) and its duration to the time over threshold (ToT).
Two types of silicon-pad sensors are under consideration, with a trade off between the signal-to-noise ratio and leakage current.
The design of the front-end modules has been recently thermally optimized with an Aluminium-Nitride ceramic baseplate to maximize the lifetime of the sensors.

The development and prototyping of the FBCM is on-track for the Phase-2 CMS upgrade for HL-LHC.
FBCM will be instrumental for achieving the target precision in luminosity measurement (2\% online; <1\% offline) during the HL-LHC era.

\acknowledgments
We acknowledge the support by the following institutes and funding
agencies: CERN; 
the national research projects RVTT3 "CERN Science Consortium of Estonia" and PUT PRG1467 "CRASHLESS" (Estonia);
Helmholtz-Gemeinschaft Deutscher Forschungszentren (HGF) (Germany); National Research, Development and Innovation Office (NKFIH), including contract numbers K 143460 and TKP2021-NKTA-64 (Hungary);
the US CMS operations program, the US National Science Foundation (NSF), and the US Department of Energy (DOE) (USA).

\bibliographystyle{JHEP}
\bibliography{references}

\providecommand{\href}[2]{#2}\begingroup\raggedright\begin{thebibliography}{10}

\bibitem{BRIL-TDR}
{\scshape CMS} collaboration, \emph{{The Phase-2 Upgrade of the CMS Beam Radiation Instrumentation and Luminosity Detectors}},  Tech. Rep. \href{https://cds.cern.ch/record/2759074}{CERN-LHCC-2021-008, CMS-TDR-023}, CERN, Geneva (2021).

\bibitem{LUMI-15-16}
{\scshape CMS} collaboration, \emph{Precision luminosity measurement in proton-proton collisions at $\sqrt{s}$ = 13 {TeV} in 2015 and 2016 at {CMS}}, \href{https://doi.org/10.1140/epjc/s10052-021-09538-2}{\emph{Eur. Phys. J. C Part. Fields} {\bfseries 81} (2021) 800}.

\bibitem{EMITTANCE-SCAN}
O.~Karacheban and P.~Tsrunchev, \emph{Emittance scans for {CMS} luminosity calibration}, \href{https://doi.org/doi:10.1051/epjconf/201920104001}{\emph{EPJ Web of Conferences} {\bfseries 201} (2019) 04001}.

\bibitem{FBCM-ASIC}
J.~Kaplon et~al., \emph{The optimization, design and performance of the {FBCM23 ASIC} for the upgraded {CMS} beam monitoring system}, \href{https://doi.org/doi:10.48550/arXiv.2312.02834}{\emph{TWEPP2023 conference proceedings} (2023) }.

\bibitem{BCM1F-diamond}
{A. A. Zagozdzinska et al. on behalf of the CMS Collaboration}, \emph{New fast beam conditions monitoring {(BCM1F)} system for {CMS}}, \href{https://doi.org/doi:10.1088/1748-0221/11/01/C01088}{\emph{Proc. Topical Workshop on Electronics for Particle Physics (TWEPP15), JINST 11 (2016) C01088} (2016) }.

\bibitem{BCM1F-diamond-2}
{M. Guthoff on behalf of the CMS Collaboration}, \emph{The new fast beam condition monitor using poly-crystalline diamond sensors for luminosity measurement at {CMS}}, \href{https://doi.org/doi:10.1016/j.nima.2018.11.071}{\emph{Proc. 14th Pisa Meeting on Advanced Detectors (Pisameet), Nucl. Instrum. Meth. A 936 (2019) 717} (2018) }.

\bibitem{BCM1F-RUN3}
{J. Wańczyk on behalf of CMS Collaboration}, \emph{Upgraded {CMS} fast beam condition monitor for {LHC Run 3} online luminosity and beam induced background measurements}, \href{https://doi.org/10.18429/JACoW-IBIC2022-TH2C2}{\emph{JACoW} {\bfseries IBIC2022} (2022) 540}.

\bibitem{TRACKER-TDR}
{\scshape CMS} collaboration, \emph{{The Phase-2 Upgrade of the CMS Tracker}},  Tech. Rep. \href{https://cds.cern.ch/record/2272264}{CERN-LHCC-2017-009, CMS-TDR-014}, CERN, Geneva (2017).

\bibitem{Orfanelli:2022zhe}
{S. Orfanelli on behalf of the CMS Collaboration}, \emph{{The CMS Inner Tracker electronics system development}}, \href{https://doi.org/10.1088/1748-0221/17/08/C08003}{\emph{JINST} {\bfseries 17} (2022) C08003}.

\bibitem{Faccio:2020rae}
F.~Faccio et~al., \emph{{The bPOL12V DCDC converter for HL-LHC trackers: towards production readiness}}, \href{https://doi.org/10.22323/1.370.0070}{\emph{PoS} {\bfseries 370} (2020) 070}.

\bibitem{Biereigel:2020jri}
S.~Biereigel et~al., \emph{{The lpGBT PLL and CDR Architecture, Performance and SEE Robustness}}, \href{https://doi.org/10.22323/1.370.0034}{\emph{PoS} {\bfseries TWEPP2019} (2020) 034}.

\bibitem{Soos:2017stv}
C.~So\'os et~al., \emph{{Versatile Link PLUS transceiver development}}, \href{https://doi.org/10.1088/1748-0221/12/03/C03068}{\emph{JINST} {\bfseries 12} (2017) C03068}.

\bibitem{Apollo}
{E. S. Hazen et al. }, \emph{{The Apollo ATCA platform}}, \href{https://doi.org/doi:10.22323/1.370.0120}{\emph{PoS} {\bfseries TWEPP2019} (2020) 120}.

\bibitem{FBCM-grease}
M.~Ferrari et~al., \emph{Selection of radiation tolerant commercial greases for high-radiation areas at {CERN}: Methodology and applications}, \href{https://doi.org/doi:10.1016/j.nme.2021.101088}{\emph{Nuclear Materials and Energy} {\bfseries 29} (2021) 101088}.

\bibitem{PT1000}
V.~Miklyaev et~al., \emph{Application of {Pt1000} c420 thin-film temperature sensors at superconducting and other types of facilities}, \href{https://doi.org/doi:10.1134/S1547477120010124}{\emph{Phys. Part. Nucl. Lett. 17 (2020) 1, 44-56} (2020) }.

\end{thebibliography}\endgroup

\end{document}